# Controlling the superconducting transition by spin-orbit coupling

N. Banerjee[1,2]*, J. A. Ouassou[3,4], Y. Zhu[2], N. A. Stelmashenko[2], J. Linder[3,4+] and M. G. Blamire[2]

[1] Department of Physics, Loughborough University, Epinal Way, Loughborough, LE11 3TU, UK.

[2] Department of Materials Science and Metallurgy, University of Cambridge, 27, Charles Babbage Road, Cambridge CB3 0FS, UK.

[3] Department of Physics, Norwegian University of Science and Technology, N-7491 Trondheim, Norway.

[4] QuSpin Center of Excellence, Department of Physics, Norwegian University of Science and Technology, N-7491 Trondheim, Norway

*N.Banerjee@lboro.ac.uk

+jacob.linder@ntnu.no

Whereas there exists considerable evidence for the conversion of singlet Cooper pairs into triplet Cooper pairs in the presence of inhomogeneous magnetic fields, recent theoretical proposals have suggested an alternative way to exert control over triplet generation: intrinsic spin-orbit coupling in a homogeneous ferromagnet coupled to a superconductor. Here, we proximity-couple Nb to an asymmetric Pt/Co/Pt trilayer, which acts as an effective spin-orbit coupled ferromagnet owing to structural inversion asymmetry. Unconventional modulation of the superconducting critical temperature as a function of in-plane and out-of-plane applied magnetic fields suggests the presence of triplets that can be controlled by the magnetic orientation of a single homogeneous ferromagnet. Our studies demonstrate for the first time an active role of spin-orbit coupling in controlling the triplets – an important step towards the realization of novel superconducting spintronic devices.

Conventional superconductivity arises from an attractive pairing of spin-up and spin-down electrons, whereas ferromagnetism is due to an imbalance in the number of spin-up and spin-down electrons. In superconductor/ferromagnet (S/F) proximity structures, the competing nature of these two orders is the source of rich physics [1,2]. For example, two opposite spin-paired electrons of a Cooper pair enter different spin bands upon transmission into an adjacent F layer, resulting in a finite centre-of-mass momentum. This results in a weak oscillatory dependence of the superconducting transition temperature $T_C$ superimposed on the monotonic $T_C$ suppression due to increasing F layer thickness [3,4]. In more complex F/S/F trilayers, $T_C$ is higher when the F moments are antiparallel than parallel [1,5–7], arising from the higher net pair-breaking exchange field in the parallel state. This *spin-switch effect* allows an active control of $T_C$ using magnetic states.

In contrast, S/F'/F and F'/S/F systems have recently shown an enhancement in the proximity effect between the S and the F layers [8–11] for non-collinear F-moment alignments. This unusual proximity effect results from a conventional spin-zero singlet Cooper pair being transformed into a spin-one triplet Cooper pair. These *long-ranged triplets* (LRTs) consist of electrons from the same spin band, and are therefore immune to the pair-breaking exchange field in F, thereby enhancing the coupling between the two layers. The increased coupling makes superconductivity spread across the whole system, which reduces $T_C$ by up to 120-400 mK [8,10,12]. Although the control of superconductivity by modulating magnetic states is attractive for applications in superconducting spintronics [13–17], precisely controlling the relative angle between the magnetic layers is difficult [8–10,17–19]. Practical applications require a simplified structure with fewer interfaces to minimize spin-scattering, motivating the exploration of alternative mechanisms for triplet generation.

Theoretical studies [20–27] predict that spin-orbit coupling (SOC) in S/F bilayers can produce an anisotropic depairing effect on triplets. The Cooper pair spin direction being determined by the F layer moment, then implies that in an S/F bilayer with SOC, triplets can be controlled by the magnetisation direction of a *single* homogeneous F [25]. This Letter reports measurements on Nb/Pt(*x*)/Co/Pt structures where the structural inversion asymmetry gives rise to a SOC for *x*>0 [28,29]. We compare the $T_C(H)$ behaviour between samples with and without SOC to demonstrate the role of a triplet proximity effect in the former, and confirm the prediction that $T_C$ can be controlled by rotating a single homogeneous magnetic layer in SOC systems.

The thin-film stacks were deposited by dc magnetron sputtering in an ultrahigh vacuum chamber onto unheated oxidized silicon (100) substrates placed on a rotating table. The substrates passed under magnetrons whose power, and the rotation speed of the substrate table, were adjusted to control the layer thicknesses (thicknesses in nanometers in parentheses). The Pt and Co layer thicknesses were adjusted to tune the in-plane (IP) and out-of-plane (OOP) magnetic anisotropy, allowing control over the angle between the magnetization and the sample plane by applying moderate magnetic fields, and so control the effectiveness of the singlet/triplet conversion. During deposition, the chamber was cooled by a liquid nitrogen jacket to achieve a pressure below 3 × $10^{-7}$ Pa. The layers were sputtered in 1.5 Pa Ar. Control samples of Nb/Pt and Nb/Co/Pt and samples with varying Pt and Nb thickness were also deposited.

Figure 1(a,b) shows the magnetization *M* vs. applied field *H* for Nb(24)/Co(1.5)/Pt(1.5) and Nb(24)/Pt(2)/Co(1.5)/Pt(1.5) samples measured at 10 K using a SQUID magnetometer. The blue (red) points, represent the magnetic field applied IP (OOP). While for the Nb/Co/Pt stack the magnetization preferentially lies IP [Fig. 1(a)], insertion of a 2 nm Pt layer at the Nb/Co interface [Fig. 1(b)] results in a clear hysteretic switching for IP and OOP applied fields. This allows us to control the magnetization tilt with respect to the film plane using moderate magnetic fields. Perpendicular magnetic anisotropy in Pt/Co systems [30,31] is generally attributed to an enhancement in the perpendicular Co orbital moment resulting from a Pt 5*d* – Co 3*d* hybridization. The OOP anisotropy is inversely proportional to the Co layer thickness [32] and here a 1.5 nm Co allows us to control the tilt using low magnetic fields.

Transport measurements were performed on unpatterned samples in the range of 3-8 K using four-point resistance measurement technique in a pulsed-tube cryocooler. A constant bias current of 5 μA was used. The magnetic field was applied by ramping it up in steps of 5-10 mT from zero-field, and each $T_C$ measurement was carried out in constant field. The maximum width of the $T_C$ ~ 180 mK.

Figure 2 shows $T_C(H)$ for the three different samples. For most of the samples there is an apparent difference between the zero-field $T_C$ for IP and OOP measurements likely arising due to the different relative positions of the sample holders with respect to the temperature sensor. Several measurements from the same sample shows that this difference is random and magnetic field-independent, without affecting the overall trends of the $T_C$ dependence. This possibly arises due to minor differences in steady-state gas flow conditions between each cooling cycle.

In Figs. 2(a,b) for a Nb(24)/Pt(2) bilayer and Nb(24)/Co(1.5)/Pt(1.5) trilayer with IP magnetic field, the $T_C$ remains roughly constant up to 120 mT with random fluctuations of less than 15 mK. Figure 2(c) shows measurements for Nb(24)/Pt(2)/Co(1.5)/Pt(1.5) stack. Strikingly, the data show a clear, rapid $T_C$ suppression of 40 mK from 0–100 mT. The full data range for all three samples (Supplemental Material Fig. 4) shows that the $T_C$ suppression below 100 mT for Nb(24)/Pt(2)/Co(1.5)/Pt(1.5) is comparable to the $T_C$ suppression for the other two structures for the entire field range up to 500 mT. While the net constant field-induced $T_c$ suppression of ~60 mK till 500 mT for all the structures can be explained by a weak field-induced depairing for 24 nm thick Nb films, the explanation for the $T_C$ suppression by 40 mK at low fields for the structure with an additional Pt interlayer is not straightforward. From the systematic layer sequences, it is clear that the extra Pt layer between Nb and Co plays a role. For the OOP fields [Figs. 2(d-f)], all samples show a pronounced $T_C$ suppression due to the strong orbital depairing from the applied field.

Before attempting to explain our results in terms of SOC-induced control of triplets, we quantify the flux-induced $T_C$ modulation which arises from the OOP magnetization of Co-containing samples [Figs. 2(e,f)]; for the Nb/Pt sample there is no magnetic moment, so the suppression shown in Fig 2(d) must originate purely from the orbital depairing effect. The Co layer in the Nb/Co/Pt sample has an IP anisotropy with an OOP saturation field of ~120 mT [Fig 1(a)]; the corresponding $T_C(H)$ plot [Fig. 2(e)] shows a rapid $T_C$ suppression in the field range below this value, which can be partially explained by the magnetization-induced flux density being drawn OOP and adding to the applied field. A similar effect would be expected for the Nb/Pt/Co/Pt sample, albeit with a lower saturation field reflecting the OOP anisotropy [Fig. 1(b)]; in fact, the low-field suppression of $T_C$ is *lower* than that for the Nb/Co/Pt sample implying that a different, partially compensating, $T_C$-modulating effect must be at work. This behaviour is more pronounced for Nb/Pt/Co/Pt containing a thinner 18 nm Nb (Fig. 2f inset). The OOP $T_C$ suppression is expected to be significantly more considering the enhanced orbital depairing in thinner Nb films (Supplemental Material section VII). Furthermore, domain-wall induced suppression of superconductivity can be ruled out since at higher fields elimination of domain walls should restore superconductivity. This is in sharp contrast to Fig. 2c, where superconductivity is suppressed at larger IP fields.

The role of an unconventional proximity effect in the Nb/Pt/Co/Pt sample is further strengthened by the IP $T_C$ data in Figs. 2(a-c). The data in Fig. 2(a) without a magnetic layer demonstrate small orbital depairing in the IP configuration with a $T_C$ change <15 mK below 120 mT. Similar behavior is observed for the Nb/Co/Pt sample [Fig. 2(b)], for which the IP anisotropy means that an IP field does not modify the moment. In contrast, the Nb/Pt/Co/Pt sample shows a decrease in $T_C$ of 50 mK in the same range; if this $T_C$ modulation arose from field-induced changes to the flux injection from the Co layer, the $T_C$ would have been expected to have *increased* due to a reduction in the OOP component of magnetization. Similar behaviour is observed for thinner Nb: $T_C$ remains roughly constant at low IP fields for a Nb(20)/Pt(2) bilayer (fluctuations <10 mK), but is suppressed by 90 mK for Nb(18)/Pt(2)/Co(1.5)/Pt(1.5) (Fig. 2c inset). $T_C$ changes arising from anisotropic interface magnetoresistance in Pt/Co/Pt structures [33,34] can be ruled out as the resistance changes would be an order of magnitude smaller than here.

To summarise: while the Nb/Pt and Nb/Co/Pt results can be qualitatively explained in terms of flux and field-induced orbital depairing, the Nb/Pt/Co/Pt behavior is distinctly different, and a rapid low-field $T_C$ suppression is induced for IP field which tends to align the Co magnetisation parallel to the Nb plane.

The key to understanding our results is that the proximity effect in S/F systems with a single homogeneous F layer cannot be controlled by changing the magnetisation angle with respect to the film plane (after subtracting the effect of flux injection from the F layer). In S/F'/F systems, non-collinear F and F' layer moments generate LRTs, which enhance the proximity coupling between S and F and so decreases $T_C$. However, in presence of SOC, the picture dramatically changes even for a S/F system, and the energy of the triplets depends on the magnetisation angle with the film plane. Here, the SOC arises in the form of a Rashba coupling $\alpha$ in the inversion asymmetric Pt/Co/Pt trilayer. Using the Usadel equation for an S/F system with SOC it can be shown that, while the singlet Cooper pair effective energy $E_s = \epsilon$ is unaffected by the SOC, the triplet Cooper pairs get effective energies $E_t(\theta) = \epsilon + iD\alpha^2[3 - \cos(2\theta)]$, where $\theta$ is the angle that the triplet ***d***-vector makes with the thin-film plane. The ***d***-vector quantifies the triplet spin direction [19], and is roughly parallel to the exchange field (Supplemental Material). The effective triplet energy being imaginary can be interpreted as a depairing term, implying a stronger suppression of triplets with an OOP than IP orientation. Rotating the magnetic field from an IP to OOP orientation increases the depairing term from $2iD\alpha^2$ to $4iD\alpha^2$, resulting in a stronger suppression of triplet superconductivity. Since closing the triplet proximity channel results in fewer Cooper pairs leaking out of the superconductor, this increases the $T_c$ (Supplemental Material).

We have modelled our results using the Usadel equations for a diffusive system with intrinsic SOC. We assume a coherence length of 15 nm for the Nb (5 nm mean free path) [35,36] and treat the Pt/Co/Pt trilayer as a single effective ferromagnet with SOC. This approach has two advantages: firstly, scattering at Pt/Co interfaces allows us to use a diffusive model without microscopic details of the interface; secondly, the exchange splitting of the Co layer is now averaged out over a larger volume allowing us to use a quasiclassical approach. We have assumed an average exchange splitting of 140 meV for Pt/Co/Pt based on previously reported values of 300 meV for Co [37]. The S/F interfaces were modelled using the Kupriyanov-Lukichev boundary conditions (tunnelling limit), and the tunnelling conductance providing the best estimate of the $T_C$ for the structure at zero field was used. Following this, we obtained a tunnelling conductance of 65% for the Nb/[Co/Pt] and 85% for the Nb/[Pt/Co/Pt] interfaces, relative to the bulk conductance of each material. The Rashba coupling strength was determined as follows: since the $T_C(H)$ variation arises from the Rashba SOC, the experimental ratio $T_C$(H=120 mT)/ $T_C$(H=0 mT) was calculated, and the Rashba coupling varied to match this ratio. This depended on the relative IP and OOP magnetization components, and we selected the best fit 6.5×$10^{-11}$ eVm. This value is close to ~5×$10^{-12}$ eVm for asymmetric Pt/Co/Pt structures estimated from [38]. The higher values in our system could arise due to different Pt and Co thicknesses and interfaces, which strongly influence the Rashba coupling [29]. The only free parameter in our model was the angle between the magnetization and film plane ($\theta$) at zero field, which we assumed to be 45° from a best fit to the experimental data. The consistency is seen by extrapolating the magnetization data to zero field for IP and OOP orientations [Fig. 1(b)]. The detailed magnetization model is given in the Supplemental Material, but importantly the initial magnetization angle assumed did not change the trend of $T_C$ variation with applied field, but only altered the magnitude of the $T_C$ shift in the IP orientation within reasonable limits.

In the Nb/Co system without SOC, the $T_C$ modulation for OOP fields would, purely from the orbital depairing, give rise to a monotonic decrease with increasing magnetic field. Similarly, without SOC, for IP fields $T_C$ would remain unchanged assuming zero orbital depairing. We have numerically calculated the $T_C$ vs IP and OOP field for Nb(24)/Co(1.5)/Pt(1.5) and Nb(24)/Pt(2)/Co(1.5)/Pt(1.5) obtained by solving the Usadel equations and incorporating a Rashba-type SOC for the Nb/Pt/Co/Pt structure (Supplemental Material Figs. 1 and 2). It is immediately clear that by incorporating a SOC term, $T_c$ depends on $\theta$ (Fig. 2), whereas it is independent of the angle without SOC (Fig. 1).

We plotted the difference $\Delta T_C = T_C^{\mathrm{Nb/Pt/Co/Pt}} - T_C^{\mathrm{Nb/Co/Pt}}$ between the results for Nb(24)/Pt(2)/Co(1.5)/Pt(1.5) and Nb(24)/Co(1.5)/Pt(1.5) in Fig.3 and, for easier comparison with theory, shifted the $\Delta T_C$ origin to zero. The figure shows the numerically calculated $\Delta T_C$ for IP and OOP applied fields (a,b), and the corresponding experimental results (c,d). The variation can be understood from the discussion above where the $T_C$ suppression for OOP fields for Nb(24)/Co(1.5)/Pt(1.5) is purely due to orbital effects, whereas in Nb(24)/Pt(2)/Co(1.5)/Pt(1.5) there is a compensating effect due to suppression of the spin-zero triplet (short-ranged triplets, SRT) generation resulting in an increasing $\Delta T_C$ with the applied field. For IP fields with negligible orbital depairing, $\Delta T_C$ decreases due to an enhancement of the proximity effect in Nb(24)/Pt(2)/Co(1.5)/Pt(1.5) arising from an increased SRT generation. The overall numerical trend is in excellent agreement with the experiments as shown by comparing Figs. 3(a,c) (IP fields) and 3(b,d) (OOP fields) for Nb(24)/Pt(2)/Co(1.5)/Pt(1.5) structure. The magnitude of $\Delta T_C$ from the simulation (~22mK) is 55% of that of the experimental value of 40 mK. The lower simulated $\Delta T_C$ values can arise due to a simplified model where we have assumed an ideal interface and a simplified magnetic model. In real systems, interdiffusion and interface roughness can affect the magnitude of $\Delta T_C$.

To better understand the role of the Pt layer at the Nb/Co interface, we have examined $T_C(H)$ variation with the thickness of this layer. Figure 4 shows $\Delta T_C$ for Nb(24)/Pt(x)/Co(1.5)/Pt(1.5) and Nb(24)/Co(1.5)/Pt(1.5) for x= 0.3 and 1. The top (bottom) panels show the $\Delta T_C$ for IP (OOP) fields. While for IP fields $\Delta T_C$ is ~ 15 mK change for 0.3 nm (Fig. 4a), the 1 nm structure shows $\Delta T_C$ ~ 25 mK drop superimposed on the noise (Fig. 4b). Comparison of the *M*(*H*) loops (Supplemental Material Fig. 5) shows a transition from fully IP to a clear hysteretic switching for both IP and OOP magnetization with increasing bottom Pt thickness from 0.3 nm to 2 nm. The corresponding IP field-dependent $T_C$ for 2 nm Pt shows a large change ~50 mK at low fields (Figs. 2c and 3c). The large change in low-field $T_c$ appears only in the region where the IP magnetization approaches saturation, beyond which the $T_c$ suppression is comparable for all the structures. This indicates the active role played by the magnetization angle in modulating $T_C$ for structures showing a comparable IP and OOP anisotropy in addition to the presence of a Pt layer at the Nb/Co interface. $\Delta T_C$ for the OOP field [Figs. 4(d,e)] increases with applied field and similar to the IP $\Delta T_C$, the magnitude of this change increases with thicker Pt layer at the Nb/Co interface.

The role of the interfacial Pt layer is most strikingly evident in the normalised $\Delta T_C$ between Nb/Co/Pt and Nb/Pt [Figs. 4 (c,f)]. The IP $\Delta T_C$ (Fig. 4c) remains constant (~10 mK fluctuation) and the OOP $\Delta T_C$ decreases with increasing applied field, in sharp contrast with Nb/Pt/Co/Pt structures. This can easily be explained by equal negligible orbital depairing for IP fields in both structures and increased flux injection in Nb/Co/Pt for OOP fields which suppresses the $T_C$ more rapidly for Nb/Co/Pt. Simulations similar to Fig. 3 for 0.3 and 1 nm Pt thickness samples support the experimental trend seen in Fig. 4 (Supplemental Material, Fig. 3).

Our measurements possibly underestimate the magnitude of the SOC-induced $T_C$ change. This is because the increased OOP magnetisation with increasing *x* in Nb/Pt(*x*)/Co/Pt results in more Co flux being directed into Nb. This reduces $T_C$ as *x* is increased, which can counteract some of the $T_C$ increase caused by the stronger SOC associated with increasing *x.* This implies that even though we see a finite non-zero $\Delta T_C$ for OOP fields for x=0.3 and x=1.0, the actual SOC-induced changes get progressively higher with increasing Pt thickness to compensate for the increasing flux injection from OOP magnetisation.

SOC introduces two competing effects – triplet depairing due to imaginary terms in the effective energy, and LRT generation due to triplet mixing terms [25]. Numerically we found the energy-penalty of the SRT is more important than the LRT generation for the $T_C$ modulation. We reiterate an important point: SOC couples the magnetization with the SRT energy, which is different to spin-relaxation effects induced by SOC on superconductivity [35]. In S/F structure without SOC, the SRT energy is *independent* of the magnetization state, and $T_C$ is independent of the magnetization angle $\theta$. However, in presence of SOC the SRT energy depends on $\theta$; with an increasing OOP field, the 'leakage' of the Cooper pairs through the triplet channel is reduced, thereby increasing the $T_C$ (since the superconducting gap directly depends on the singlet pair amplitude). As the magnetization is made IP, the SRT generation is energetically more favourable, thereby 'draining' the superconductor of Cooper pairs and reducing $T_C$. There is thus a qualitative difference between the samples for which SOC is expected to be relevant and those which 'simply' have a magnetic layer whose magnetic orientation controls the injected flux.

The results reported here cannot be explained by conventional S/F proximity theory without considering SOC. While the superconducting spin-valve with a single homogeneous ferromagnet demonstrated here drastically simplifies the control of superconductivity, a natural progression involves structures with combined Rashba and Dresselhaus coupling predicted to control LRT [25]. Incorporating such structures in Josephson junctions would allow the design of devices currently under intense focus in superspintronics.

Acknowledgments

This work was supported by ERC AdG 'Superspin' and EPSRC Programme Grant EP/N017242/1. N.B. acknowledge funding from Loughborough University and the British Council. J.L. and J.A.O. acknowledge funding via the Outstanding Academic Fellows program at NTNU and the Research Council of Norway Grant numbers 216700 and 240806. Funding for the Center of Excellence QuSpin from the Research Council of Norway and NTNU is gratefully acknowledged.

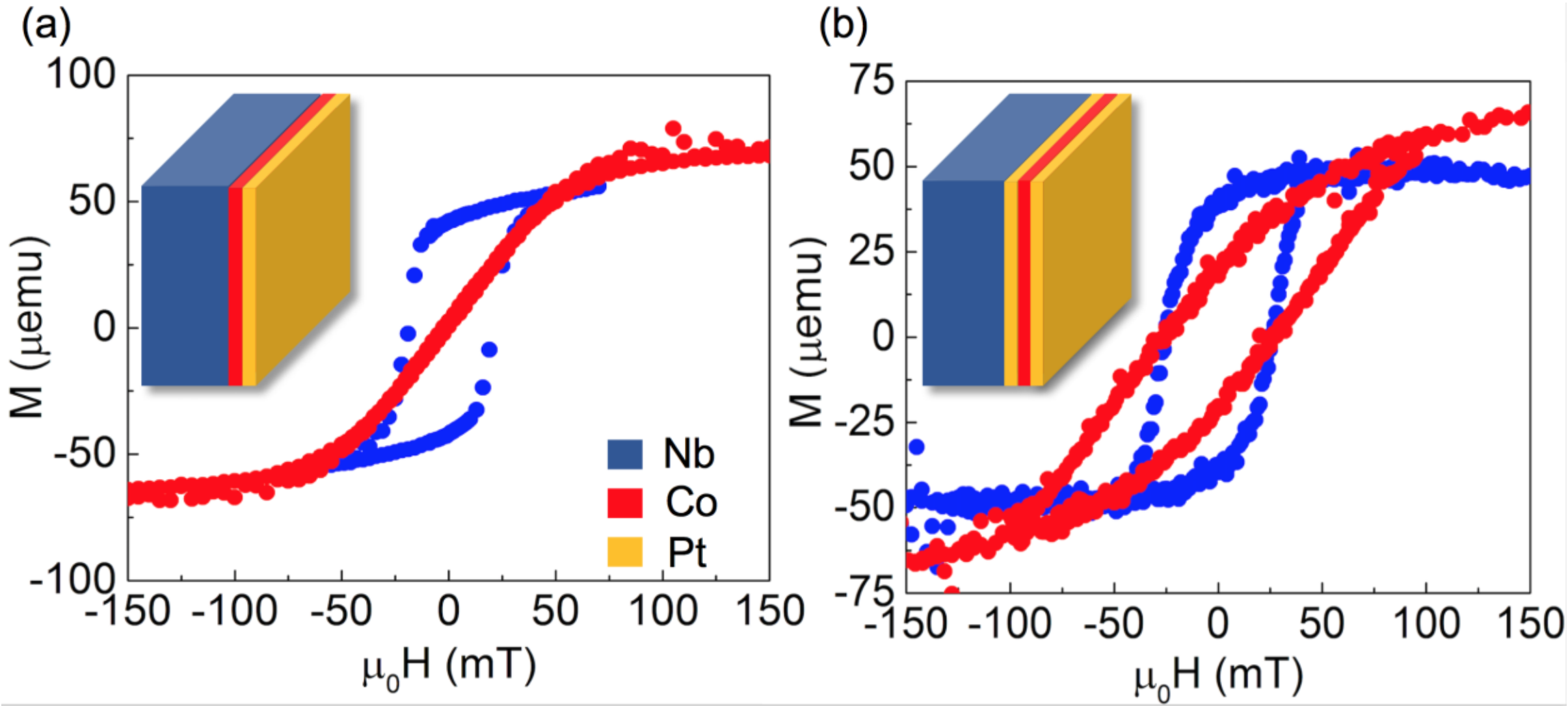

FIG. 1. (a) Magnetisation $M$ vs. applied field $H$ for a Nb(24)/Co(1.5)/Pt(1.5) structure measured at 10K, with the blue (red) points showing the field applied IP (OOP). (b) Similar measurement for an Nb(24)/Pt(2)/Co(1.5)/Pt(1.5) structure. The insets show the stack sequence.

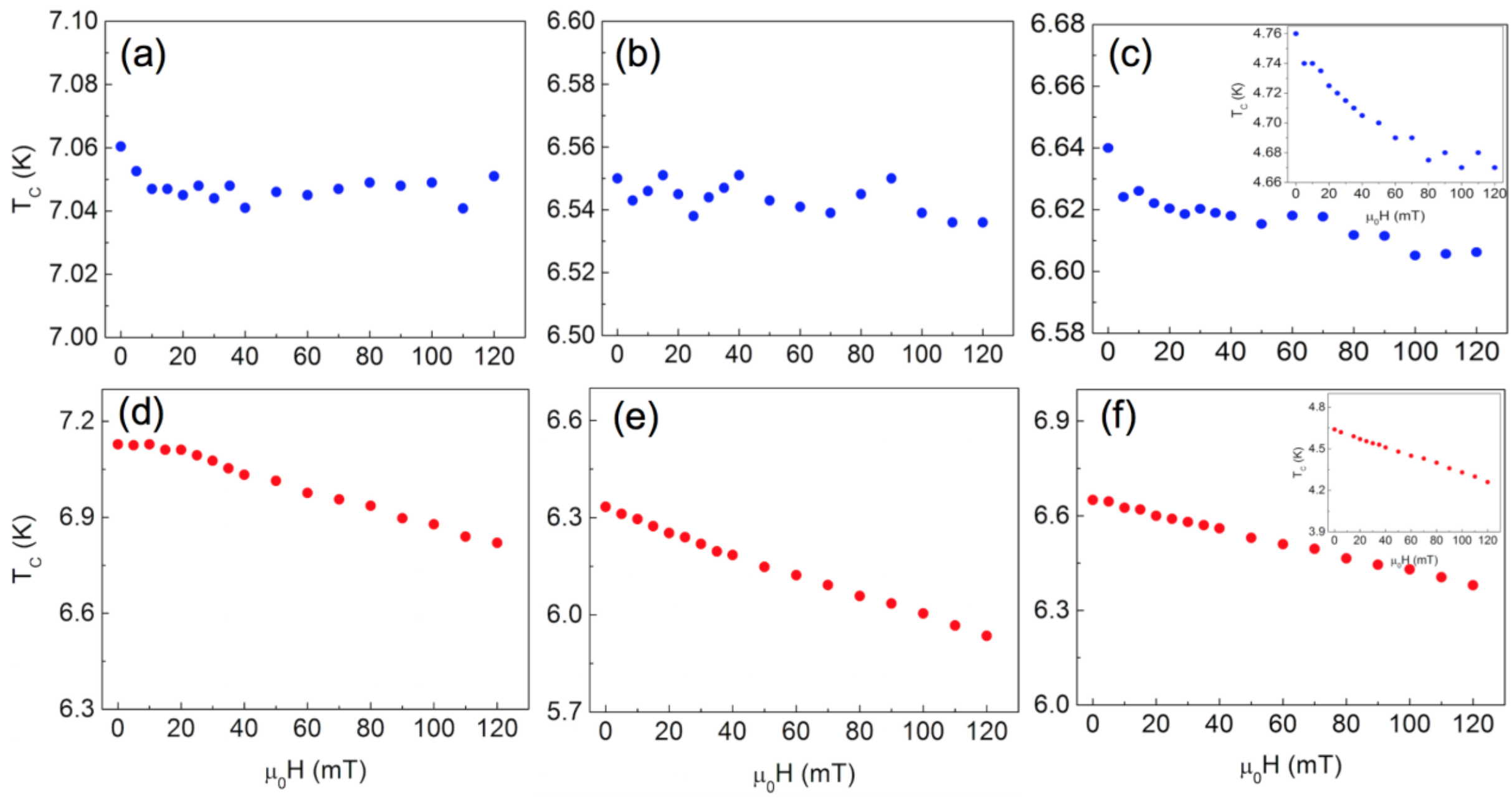

FIG. 2. Superconducting transition temperature $T_C$ plotted against the applied field $H$ for Nb(24)/Pt(2) (a,d), Nb(24)/Co(1.5)/Pt(1.5) (b,e), and Nb(24)/Pt(2)/Co(1.5)/Pt(1.5) (c,f). The rows correspond to IP (a-c) and OOP (d-f) applied fields. The insets in (c,f) shows the $T_C$ *vs* $H$ plot for Nb(18)/Pt(2)/Co(1.5)/Pt(1.5) having a thinner Nb layer with a larger $T_C$ drop ~ 90 mK for IP field.

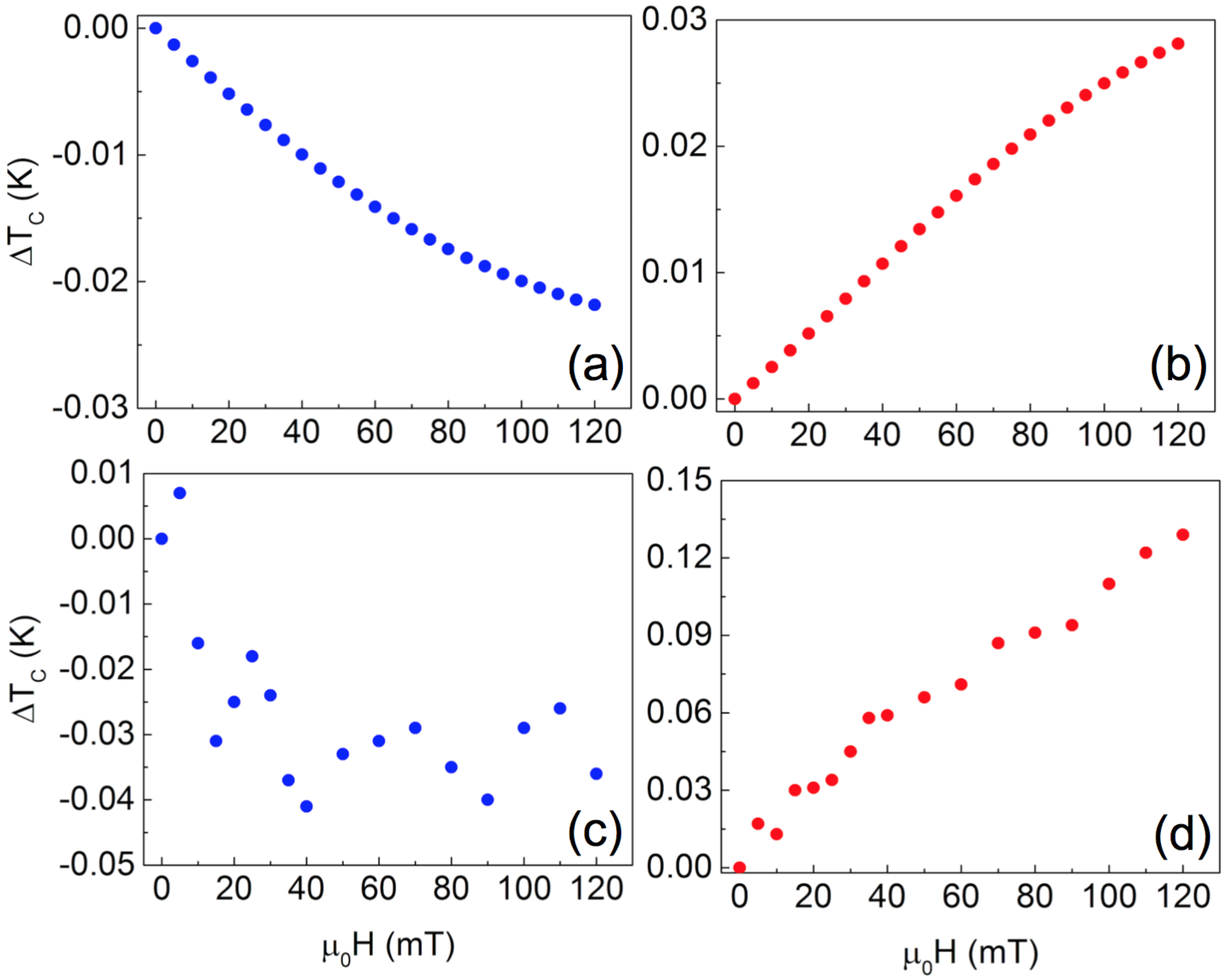

FIG. 3. Numerically calculated difference $\Delta T_C$ between the critical temperature of Nb(24)/Pt(2)/Co(1.5)/Pt(1.5) and Nb(24)/Co(1.5)/Pt(1.5) for (a) IP and (b) OOP applied fields. The corresponding experimental results for IP (c) and OOP (d) fields. The $\Delta T_C$ origin has been shifted to zero for each data set by subtracting $\Delta T_C$ ($H=0$).

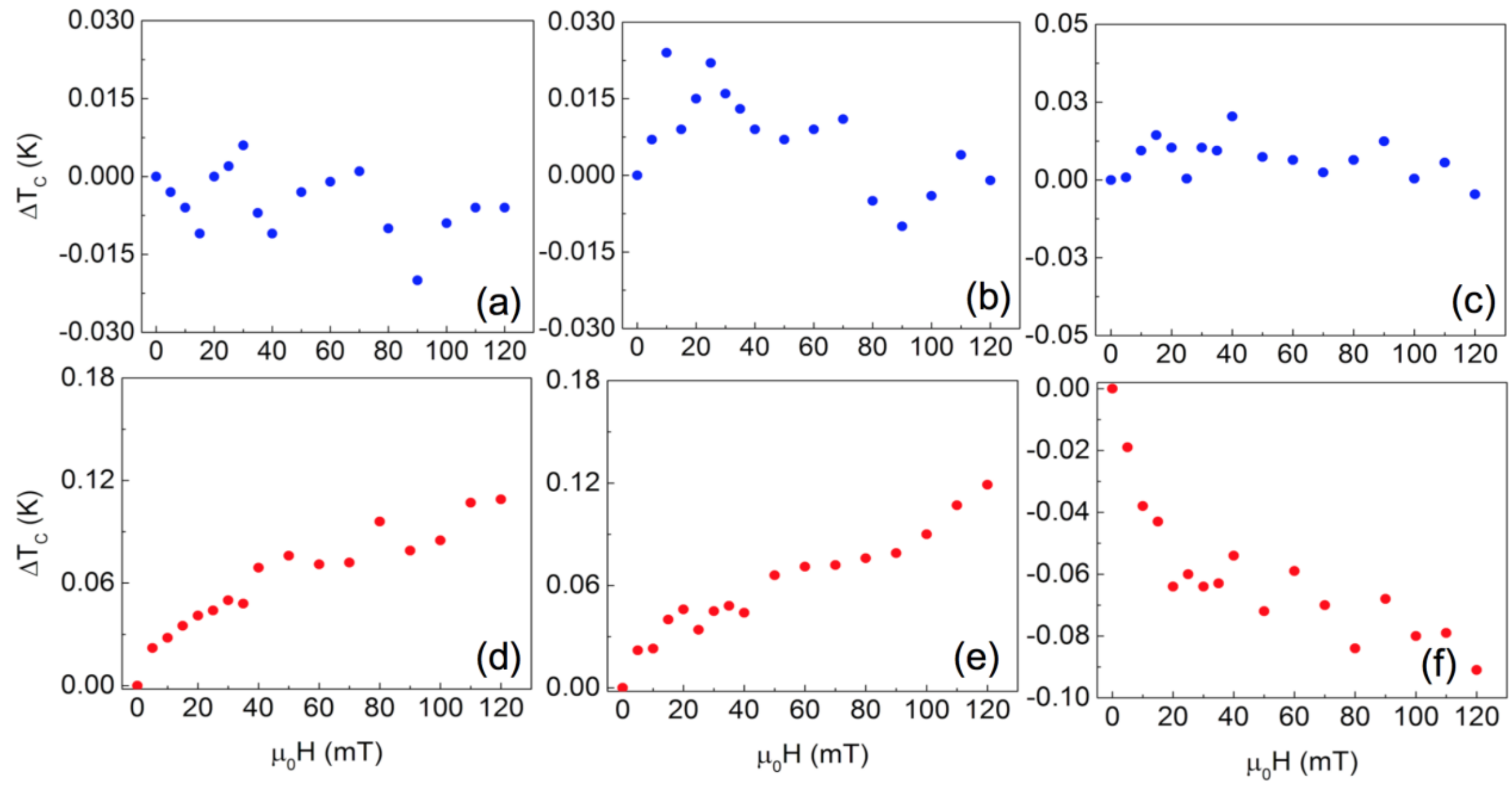

FIG. 4. Critical temperature difference $\Delta T_C$ between Nb(24)/Pt(x)/Co(1.5)/Pt(1.5) and Nb(24)/Co(1.5)/Pt(1.5) plotted against the applied fields (a,d) for IP and OOP fields with x=0.3, (b,e) IP and OOP fields with x=1.0. (c,f) $\Delta T_C$ between control samples Nb(24)/Co(1.5)/Pt(1.5) and Nb(24)/ Pt(1.5) for IP and OOP fields.

# Supplementary information: Controlling the superconducting transition by spin-orbit coupling

N. Banerjee, J. A. Ouassou, Y. Zhu, N. A. Stelmashenko, J. Linder, and M. G. Blamire

We describe the theoretical details of how the Nb/[Pt]/Co/Pt heterostructures discussed in the main manuscript can be modeled as S/F bilayers with spin-orbit coupling, and how the resulting device can function as a spin-valve. Specifically, we show how this conclusion follows from the linearized diffusion equations, describe how the full nonlinear equations were solved numerically, and finally display the results of the numerical simulations. For completeness, we also include raw experimental data for the critical temperature for applied fields up to 500 mT.

## I. INTRODUCTION

In the main manuscript, we describe the experimental setup that has been studied: Nb/[Pt]/Co/Pt structures, where Nb is a conventional $s$-wave superconductor, Co is a ferromagnetic metal, and the Pt interlayer induces a Rashba coupling in the system by breaking the inversion symmetry along the junction direction, which we take to be along the $z$-axis. In particular, we focus on the structures Nb(24)/Co(1.5)/Pt(1.5) and Nb(24)/Pt(2.0)/Co(1.5)/Pt(1.5), where the numbers in parentheses denote material thicknesses in nanometers. Both structures were modelled theoretically as S/F bilayers, where the [Pt]/Co/Pt heterostructures are treated as effective ferromagnets with thicknesses 3 nm and 5 nm, respectively. Within these effective ferromagnets, we approximate the exchange field and spin-orbit coupling as homogeneous background fields.

In Section II, we describe how the physical behaviour of these heterostructures can be understood analytically by considering the linearized diffusion equations for the system. We then describe how the full nonlinear diffusion equations were solved numerically in Section III. The magnetization model is discussed in Section IV, and the numerical results presented in Section V. In Section VI, we then describe how the parameters used for the numerical simulations were determined. Finally, we include supplemental experimental data in Section VII.

## II. ANALYTICAL TREATMENT

In the quasiclassical and diffusive limits, superconductivity is well-described by the so-called Usadel diffusion equation.[1] Near the critical temperature $T_c$, the superconducting pair amplitudes go to zero, meaning that the diffusion equation can be linearized with respect to pair amplitudes near this transition. In materials with superconductivity, ferromagnetism, and spin-orbit coupling, the linearized diffusion equations are:[2]

$$iD\,\nabla^2 f_s = \epsilon f_s + \boldsymbol{h}\boldsymbol{f}_t - \Delta, \tag{1}$$

$$iD\,\nabla^2 \boldsymbol{f}_t = \epsilon \boldsymbol{f}_t + \boldsymbol{h} f_s + 2iD\,\boldsymbol{\Omega}\boldsymbol{f}_t, \tag{2}$$

where $f_s$ is the singlet pair amplitude, $\boldsymbol{f}_t$ the triplet pair amplitude, $D$ the diffusion coefficient, $\epsilon$ the quasiparticle energy, $\Delta$ the superconducting gap, $\boldsymbol{h}$ the exchange field, and $\boldsymbol{\Omega}$ is a $3\times3$ matrix that describes the effects of the spin-orbit coupling.

From these equations alone, we can understand a lot about how the system behaves. When the superconducting gap $\Delta$ is nonzero, Eq. (1) implies that there has to be singlet pairs $f_s$ in the system as well. This makes sense, since it is precisely these singlet pairs that form the superconducting condensate of a conventional superconductor like Nb in the first place. Next, in the presence of an exchange field $\boldsymbol{h}$, some of these singlets $f_s$ are converted into triplets $\boldsymbol{f}_t$ according to Eq. (2). Note that the direction of the triplet vector $\boldsymbol{f}_t$ parametrizes the spins of the pair, and $\boldsymbol{f}_t$ is proportional to the conventional $\boldsymbol{d}$-vector.[3] The triplets generated here are oriented along the exchange field ($\boldsymbol{f}_t \parallel \boldsymbol{h}$), and are known as *short-ranged triplets* in the literature since they are exposed to the pair-breaking effects of the exchange field. Finally, Eq. (2) show that the triplet pairs are then affected by the spin-orbit matrix $\boldsymbol{\Omega}$. Depending on the structure of this matrix, the triplet pairs can either be rotated into so-called *long-ranged triplets*, or just be suppressed by the pair-breaking effect of the spin-orbit coupling.

For a Rashba coupling in the $xy$-plane – i.e. broken inversion symmetry in the $z$-direction – the matrix $\boldsymbol{\Omega}$ becomes diagonal:[2]

$$\boldsymbol{\Omega} = \alpha^2 \begin{pmatrix} 1 & 0 & 0 \\ 0 & 1 & 0 \\ 0 & 0 & 2 \end{pmatrix}, \tag{3}$$

where $\alpha$ is the Rashba coefficient. The fact that this matrix is diagonal implies that the spin-orbit coupling does not facilitate any conversion between short-ranged and long-ranged triplets. Note that this is different from the case of both Rashba and Dresselhaus coupling, and also different from the nonlinear equations (required when $T \ll T_c$). The only effect of the spin-orbit coupling here, is to change the effective energies of the IP triplets $f_x, f_y$ by $2iD\alpha^2$, and for OOP triplets $f_z$ by $4iD\alpha^2$. This energy penalty is twice as large for OOP than IP triplets, and since the triplets are again oriented along the exchange field $\boldsymbol{h}$, we note that the triplet energy penalty can effectively be adjusted by rotating the exchange field.

To make this manifest, let us parametrize the exchange field

$$\boldsymbol{h} = h\,(\cos\theta\,\boldsymbol{e}_x + \sin\theta\,\boldsymbol{e}_z), \tag{4}$$

with $\theta$ being a parameter that rotates the field from IP to OOP. We can then project Eqs. (1) and (2) along the exchange field, obtaining the scalar diffusion equations

$$iD\,\nabla^2 f_s = \epsilon f_s + h f_t - \Delta, \tag{5}$$

$$iD\,\nabla^2 f_t = E_t f_t + h f_s, \tag{6}$$

where we have defined the effective triplet energy

$$E_t(\theta) = \epsilon + iD\alpha^2(3 - \cos 2\theta)\,. \tag{7}$$

This effective energy rotates between $\epsilon + 2iD\alpha^2$ and $\epsilon + 4iD\alpha^2$ depending on the magnetization angle $\theta$. But we again note that the origin of this magnetic field dependence is that the spin-orbit coupling suppresses triplets oriented OOP more than triplets oriented IP; the magnetic field dependence only appears because the magnetic field controls what triplets we generate.

This magnetically tunable energy penalty lies at the core of the $T_c$ control discussed in this paper. By increasing the triplet energy $E_t$, we can directly suppress the triplet amplitude in the effective ferromagnet, thus closing the triplet proximity channel. Because this implies that fewer pairs will leak out of the superconductor, the singlet amplitude in the superconductor goes up, and this restores $T_c$ to higher levels.

Note that the *spin-valve effect*, i.e. the variation of the critical temperature $T_c$ with the magnetization direction $\theta$, is not a monotonic function of the spin-orbit coupling $\alpha$. If $\alpha$ is very low, then *neither* energy penalty $2iD\alpha^2$ or $4iD\alpha^2$ is high enough to significantly suppress triplets, and $T_c$ is low for all magnetic configurations. However, if $\alpha$ is very high, then *both* energy penalties are high enough to strongly suppress triplets, and $T_c$ is high for all magnetic configurations. It is for intermediate values of $\alpha$ that the spin-valve effect is maximized.

## III. NUMERICAL TREATMENT

For the numerical calculations of the critical temperature, we solved the full nonlinear diffusion equations,[2]

$$iD\tilde{\nabla}(\hat{g}\tilde{\nabla}\hat{g}) = [\epsilon\hat{\tau}_z + \hat{\Delta} + \hat{h} + \hat{\kappa}\,,\, \hat{g}], \tag{8}$$

where $\hat{g}$ is the $4\times4$ retarded quasiclassical propagator, and $\tilde{\nabla}(\,\cdot\,) = \nabla(\,\cdot\,) - i[\hat{\boldsymbol{A}},\,\cdot\,]$ is a gauge-covariant derivative that accounts for the spin-orbit coupling in the system. The other matrices in the equation can be written

$$\hat{\tau}_z = \mathrm{diag}(+1, +1, -1, -1), \tag{9}$$
$$\hat{\Delta} = \mathrm{antidiag}(+\Delta, -\Delta, +\Delta^*, -\Delta^*), \tag{10}$$
$$\hat{h} = \mathrm{diag}(\boldsymbol{h}\boldsymbol{\sigma}, \boldsymbol{h}\boldsymbol{\sigma}^*), \tag{11}$$
$$\hat{\boldsymbol{A}} = \mathrm{diag}(\boldsymbol{A}, -\boldsymbol{A}^*), \tag{12}$$
$$\hat{\kappa} = \kappa\hat{\tau}_z\hat{g}\hat{\tau}_z. \tag{13}$$

Here, $\Delta$ is the superconducting gap, $\boldsymbol{h}$ the exchange field, $\boldsymbol{\sigma}$ the Pauli vector, $\boldsymbol{A} = \alpha(\sigma_x\boldsymbol{e}_y - \sigma_y\boldsymbol{e}_x)$ the spin-orbit field,[2] and $\kappa$ is a parameter that accounts for the orbital depairing.[4]

For the interface between the superconductor and effective ferromagnet, we used the tunneling boundary conditions[5]

$$2G_0L_{\mathrm{L}}\hat{g}_{\mathrm{L}}\partial_z\hat{g}_{\mathrm{L}} = 2G_0L_{\mathrm{R}}\hat{g}_{\mathrm{R}}\partial_z\hat{g}_{\mathrm{R}} = G_T[\hat{g}_{\mathrm{L}}\,,\,\hat{g}_{\mathrm{R}}], \tag{14}$$

where $G_0$ is the normal-state conductance of each material, $G_T$ is the tunneling conductance of the interface, $\hat{g}_{\mathrm{L,R}}$ are the propagators on the left and right side of the interface, respectively, and $L_{\mathrm{L,R}}$ are the corresponding material lengths.

In order to selfconsistently determine the superconducting properties of a hybrid structure, we not only require equations for the propagator $\hat{g}$, but also an accompanying equation for the superconducting gap $\Delta$. This equation can be written[2]

$$\Delta(z) = \lambda\int_0^{\Delta_0\cosh(1/\lambda)} \mathrm{d}\epsilon\; \mathrm{Re}[f_s(\epsilon, z)] \tanh\left(\frac{\pi}{2e^\gamma}\frac{\epsilon/\Delta_0}{T/T_{cs}}\right), \tag{15}$$

where $\lambda$ is a dimensionless coupling constant, $\Delta_0$ the zero-temperature gap of a bulk superconductor, $T_{cs}$ the critical temperature of a bulk superconductor, and $\gamma$ the Euler–Mascheroni constant. We used $\lambda = 1/5$ in our simulations, and for Nb the relevant material constants are $\Delta_0 \approx 1.4$ meV and $T_{cs} \approx 9.2$ K.

In the numerical implementation, we use a Riccati-parametrization for the propagator $\hat{g}$, and employ a kind of binary search algorithm for the calculation of the critical temperature $T_c$. For more details about this procedure, see Ref. 6.

## IV. MAGNETIZATION MODEL

The measured magnetization was found to roughly follow

$$M = M_0 + \delta M \tanh(H/H_0), \tag{16}$$

where $M$ is the magnetization component along the applied field $H$. This suggests that we model the exchange field as

$$h_x/h_0 = \cos(\theta_0) + [\cos(\theta_0 - \delta\theta) - \cos(\theta_0)]\tanh(H/H_0) \tag{17}$$

in the case of an IP applied field $H$, and

$$h_z/h_0 = \sin(\theta_0) + [\sin(\theta_0 + \delta\theta) - \sin(\theta_0)]\tanh(H/H_0) \tag{18}$$

for an OOP applied field $H$. In both cases, we have assumed that the exchange field remains in the $xz$-plane, so that the relation $h_x^2 + h_z^2 = h_0^2$ can be used to find the other component. Here, $\theta_0$ is interpreted as the angle that the exchange field direction makes with the thin-film plane in the absence of external fields, while $\delta\theta$ parametrizes the maximum exchange field rotation that can be achieved using an external field.

Based on the experimental measurements, we found the saturation parameter $H_0 \approx 100$ mT to fit the data very well, but estimating $\theta_0$ and $\delta\theta$ turned out to be more difficult. We therefore fixed the first parameter to $\theta_0 = 45°$, and varied $\delta\theta \in [25°, 35°]$ to see how the results change, since the critical temperature $T_c$ is more sensitive to variations in $\delta\theta$ than $\theta_0$. Finally, the magnitude of the exchange field $h_0 = 100\Delta_0 \approx 140$ meV. This is a bit smaller than the usual estimates for Co, but is not unreasonable as an average exchange field in the [Pt]/Co/Pt heterostructure, and it is about as high as we can go in a quasiclassical model.

Using the model and parameters discussed above, we have plotted the resulting exchange field $\boldsymbol{h}$ as a function of the applied field $H$ in Fig. 1.

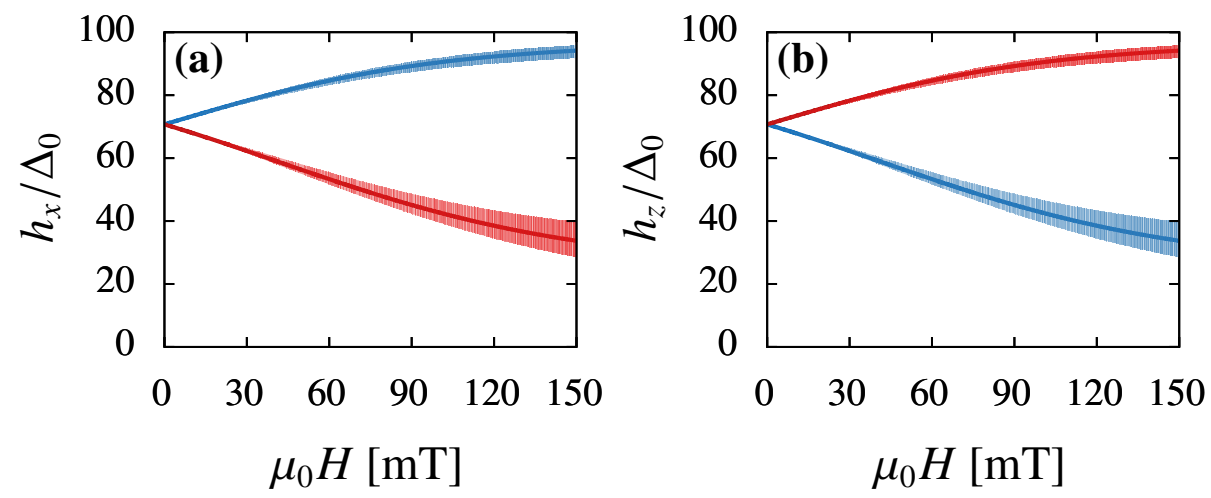

FIG. 1: Plot of the (a) IP and (b) OOP exchange field in our numerical model as functions of the applied field $H$. The blue curves correspond to an IP applied field, and the red curves an OOP applied field. The solid curves show the exchange field components for a rotation angle $\delta\theta = 30°$, while the shaded regions correspond to $\delta\theta \in [25°, 35°]$.

## V. NUMERICAL RESULTS

The results of the numerical simulations are shown in Fig. 2. Note that there is no visible shading in Fig. 2(a) because $T_c$ is only affected by orbital depairing in the absence of spin-orbit coupling, making $\delta\theta$ irrelevant. In (b) there is in fact a variation, but it is not visible in the plot because this variation is nearly two orders of magnitude smaller than the variation caused by orbital depairing. In (c), however, we do see that $\delta\theta$ does quantitatively modify the spin-valve effect. We see that the $T_c$ results are consistently $\sim 1$ K higher than the experimental results. However, the difference $\Delta T_c$ has the right shape and order of magnitude for both IP and OOP fields.

In Fig. 3, we also show the critical temperature difference $\Delta T_c$ between Nb/Pt/Co/Pt and Nb/Co/Pt for different Pt interlayer thicknesses. The results for different Pt thicknesses are more difficult to extract physical insight from. The challenge is that the thinnest Pt films do not grow uniformly in the experiment, but rather in patches, meaning that both the spin-orbit coupling $\alpha$ and interface conductance $G_T$ change in non-trivial ways at the same time. However, in all cases, we see that the experiments show a slight downward trend in $\Delta T_c$ when the sample is subjected to an IP field, and a strong upward trend when subjected to an OOP field. The effect is stronger for $x = 1.0$ and $x = 2.0$ than for $x = 0.3$, indicating that the spin-orbit coupling is likely higher in these samples due to the formation of a more uniform Pt film.

## VI. MODEL PARAMETERS

To get rid of the diffusion coefficient in Eq. (8), we used the definition $\xi = \sqrt{D/\Delta_0}$ of the diffusive coherence length, and used $\xi \approx 14$ nm for the coherence length. Using the equation $\xi \approx \sqrt{\xi_0 \ell}$, where $\xi_0 \approx 38$ nm is the ballistic coherence length of Nb and $\ell$ is the mean free path of the sample, we find that this corresponds to a reasonable mean free path $\ell \approx 5$ nm. The diffusion coefficient was assumed to the same in all materials.

Next, we estimated the tunnel conductance between the superconductor and effective ferromagnet in the junction. This was done by calculating the critical temperature $T_c/T_{cs}$ for both junctions in the absence of an external field, for various conductance ratios $G_T/G_0 \in [0, 1]$, and selecting the best possible values. We simultaneously tried to make sure that the ratio between $T_c$ for Nb/Co/Pt and Nb/Pt/Co/Pt was as close to the experimental values as possible. Unfortunately, we were unable to get a perfect quantitative fit using reasonable parameters here; but using $G_T/G_0 = 0.65$ for Nb/Co/Pt and $G_T/G_0 = 0.85$ for Nb/Pt/Co/Pt provided a qualitative match.

Note that we assume the normal-state conductance $G_0$ to be the same in Nb and the [Pt]/Co/Pt heterostructure. In reality, these two are different, and estimating an effective $G_0$ for the heterostructure from known parameters is not straight-forward. However, a difference in the normal-state conductances of the materials simply decreases the proximity-effect,[7] and the same happens if the tunneling conductance is decreased. Thus, we may compensate for a conductance asymmetry by adjusting $G_T$ accordingly—and since the tunneling conductance is already treated as a fitting parameter, this happens automatically.

Next, we discuss the orbital depairing effect. For thin-film systems, the depairing effect usually causes the critical temperature to decrease linearly with the applied field when the external field is applied OOP, and quadratically when the external field is applied IP.[8] These two cases correspond to the depairing parameters $\kappa = \Delta_0(H/H_c)$ and $\kappa = \Delta_0(H/H_c)^2$, respectively, where $H_c$ is a critical field for which $T_c \to 0$ in the absence of proximity effects. From the experimental results, we see that for an OOP case we do get a linear decrease in $T_c$ as expected. By fitting the critical temperature decay $T_c(H = 120\text{ mT})/T_c(H = 0)$ that we get from the numerical simulations to that in the Nb/Co/Pt experiment, we get an estimate $H_c \approx 1.8$ T for the critical field. For the Nb/Pt/Co/Pt structure, we simply assumed that the orbital depairing effect was the same as for Nb/Co/Pt. For the case of an IP applied field, however, we see from the experiment that the orbital depairing is negligible for $H < 150$ mT, and was therefore excluded from the IP models (i.e. $H_c = \infty$).

Finally, we estimated the Rashba coupling $\alpha \approx 12$ by fitting the ratio $T_c(H = 120\text{ mT})/T_c(H = 0)$ for the Nb/Pt/Co/Pt structures, and selecting the value of $\alpha$ that produces the best possible fits for both the IP and OOP case. This is in the unit $\hbar^2/m\xi$, where $m$ is the electron mass and $\hbar$ is Planck's reduced constant; restoring the units we get $\alpha \approx 6.5 \times 10^{-11}$ eV·m, which is very close to previous experimental estimates.

The fitting process was later repeated for Pt interlayers of thickness 0.3 nm and 1.0 nm, in place of the 2.0 nm interlayer discussed above. For the 0.3 nm case, we found a reduced Rashba coupling $\alpha \approx 9$, and the same tunneling conductance as for Nb/Co/Pt. For the 1.0 nm case, however, both these parameters were the same as for the 2.0 nm case.

## VII. EXPERIMENTAL DATA

In Fig. 4, we show how the experimentally measured critical temperature $T_c$ of various thin-film stacks behave as a function of an applied magnetic field $H$. This behaviour was recorded for fields up to the maximum limit of the equipment (500 mT). As discussed in the main paper, for low IP fields, $T_c$ remains constant for Nb(24)/Pt(2) and Nb(24)/Co(1.5)/Pt(1.5). At high fields (>120 mT), there is a small suppression ~50–70 mK. For the Nb(24)/Pt(2)/Co(1.5)/Pt(1.5) stack, there is a suppression ~50 mK even for small applied fields, which is comparable to the high field-induced suppression $\sim 60$ mK that is observed in all the structures. For OOP fields, on the other hand, there is a strong suppression in all three structures, resulting from the high orbital depairing induced by the externally applied field.

In Fig. 5, we present a comparison of the $M(H)$ loops for Nb(24)/Pt(x)/Co(1.5)/Pt(1.5) samples with $x = 0.3$ and $x = 1.0$ measured at 10 K. It is clear that for $x = 0.3$, the magnetisation preferentially lies IP with an OOP hard axis, while it starts to show a mixed anisotropy for $x = 1.0$. However, for $x = 2.0$ (fig. 1b in the main text), the sample exhibits a clear hysteretic switching, indicating a comparable IP and OOP anisotropy.

Finally, we discuss the critical temperature measurements for different Nb thicknesses. In superconducting thin films, we can use Ginzburg-Landau theory to understand the magnetic field dependence of the critical temperature $T_c$.[9] In a perpendicular field, the upper critical value $H_c(T)$ can be found from

$$\frac{\mathrm{d}H_c}{\mathrm{d}T} = -\frac{\Phi_0}{2\pi T_{c0}\xi_0^2}, \tag{19}$$

where $\xi_0$ is the zero-temperature Ginzburg-Landau coherence length, and $\Phi_0 = h/2e$ is the flux quantum. Integrating this from zero critical field ($H_c = 0$, $T = T_{c0}$) to a finite field ($H_c = H_\perp$, $T = T_c < T_{c0}$), we find that the critical temperature reduction $\Delta T_c = T_c - T_{c0}$ due to orbital depairing follows

$$\frac{\Delta T_c}{T_{c0}} = -\frac{2\pi\xi_0^2}{\Phi_0} H_\perp. \tag{20}$$

Thus, the $T_c$ variation with the field $H_\perp$ should depend only on the coherence length $\xi_0$. Fitting the observations for Nb(24)/Pt and Nb(20)/Pt, and extrapolating linearly to Nb(18)/Pt, we get an estimated $\xi_0 \approx 15.5$ nm for 18 nm Nb. We therefore expect $\Delta T_c \approx 406$ mK for Nb(18)/Pt/Co/Pt with $H_\perp = 120$ mT.

Note that the equation above ignores the additional flux injection from Co due to the magnetization rotation. To estimate a lower bound for this flux, we can rewrite Eq. (20) with the effective magnetic field $H_\perp = H_{\text{ext}} + H_{\text{int}}$, where $H_{\text{ext}}$ is the external applied field, and $H_{\text{int}}$ the internal contribution from the Co layer. Solving the resulting equation for $H_{\text{int}}$, we get

$$H_{\text{int}} = H_{\text{ext}} - \frac{\Phi_0}{2\pi\xi_0^2}\frac{\Delta T_c}{T_{c0}}. \tag{21}$$

Using the experimental $\Delta T_c$ for the Nb(24)/Co/Pt sample and $H_{\text{ext}} = 120$ mT, we estimate $H_{\text{int}} \approx 52$ mT, yielding a total field $H_\perp \approx 172$ mT. Applying an effective field $H_\perp = 172$ mT to Nb(18)/Pt/Co/Pt, we then estimate $\Delta T_c \approx 581$ mK, while the measured value was 380 mK. We note that the estimated value here only gives us a lower bound, since Nb(18)/Pt/Co/Pt is expected to have a larger flux injection from the Co layer than Nb/Co/Pt, due to the increased OOP anisotropy of the sample. A similar calculation for Nb(24)/Pt/Co/Pt gives an estimated $\Delta T_c \approx 420$ mK, while the measured value was 270 mK.

Taking the difference between the estimated and measured values above, we can attribute a critical temperature change of 201 mK to proximity effects in Nb(18)/Pt/Co/Pt, compared to 150 mK for Nb(24)/Pt/Co/Pt. Thus, we confirm that the spin-valve effect increases significantly for thinner Nb layers.

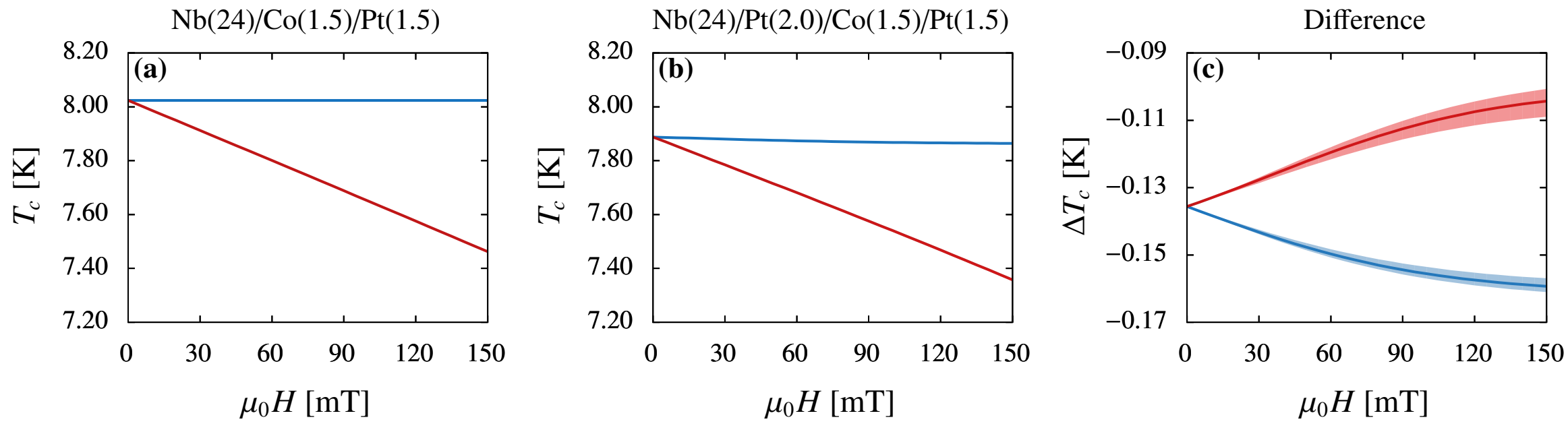

FIG. 2: Critical temperature $T_c$ as function of the applied field $H$ for (a) a model without spin-orbit coupling, (b) a model with a Rashba coupling, and (c) the difference between these results. The colors and shades have the same meanings as in Fig. 1: the blue curves correspond to an IP applied field, the red to an OOP applied field, and the shaded regions show the effect of varying the total magnetic field rotation $\delta\theta$.

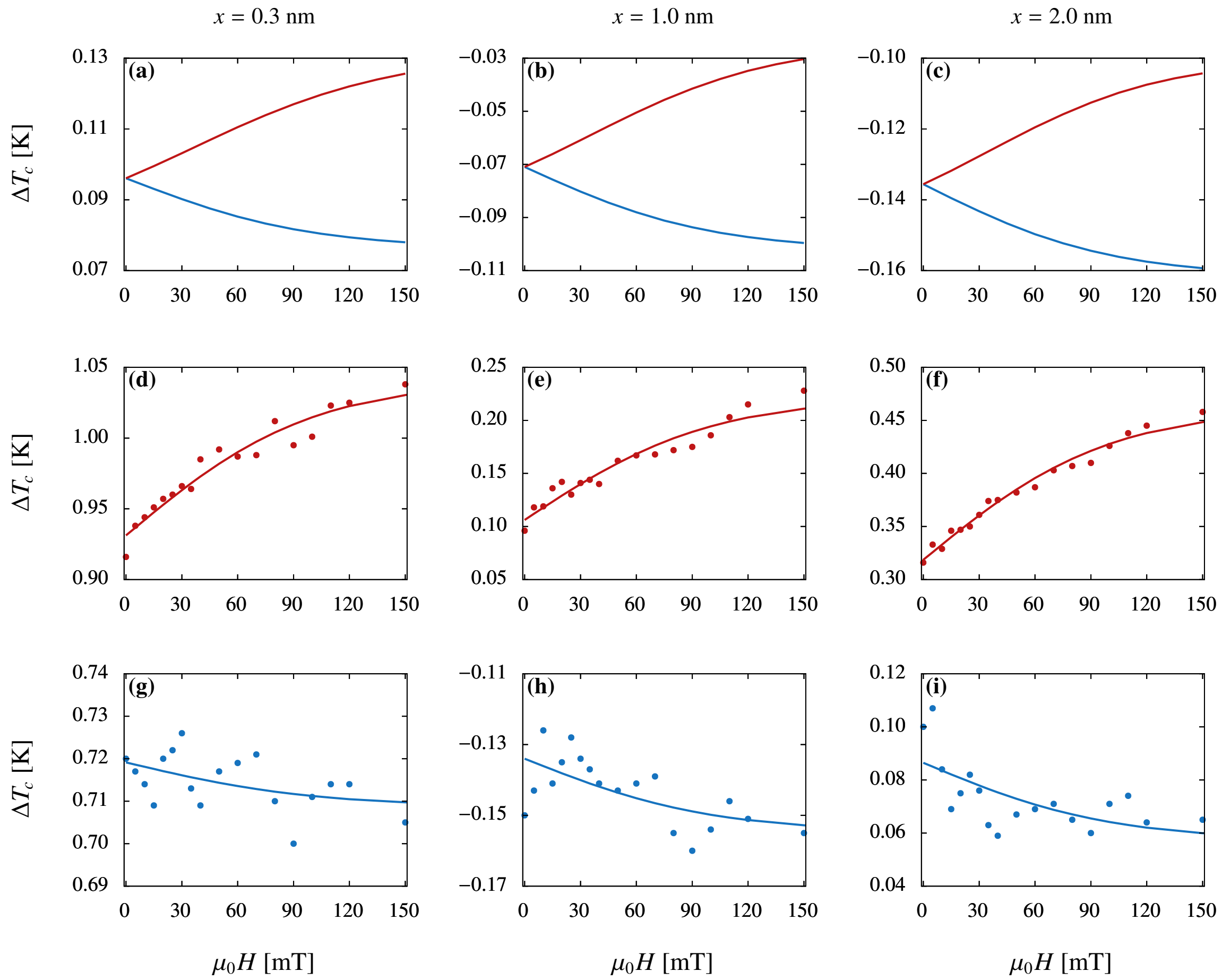

FIG. 3: Critical temperature difference $\Delta T_c = T_c$[Nb/Pt(x)/Co/Pt] $- T_c$[Nb/Co/Pt] as function of applied field $H$. The different columns correspond to different interlayer thicknesses $x$, as indicated in the column headers. The top row (a–c) shows numerical results, the middle row (d–f) shows experimental results for the OOP case, and the bottom row (g–i) shows experimental results for the IP case. As a guide to the eye, we have also plotted experimental fits to the linearization $\Delta T_c(H) = \alpha + \beta \cdot M(H)$, where the magnetization $M(H)$ is shown in Fig. 1. Note that there is some variation in the zero-field result $\Delta T_c(0)$, due to different locations of the temperature sensor with respect to the sample positions.

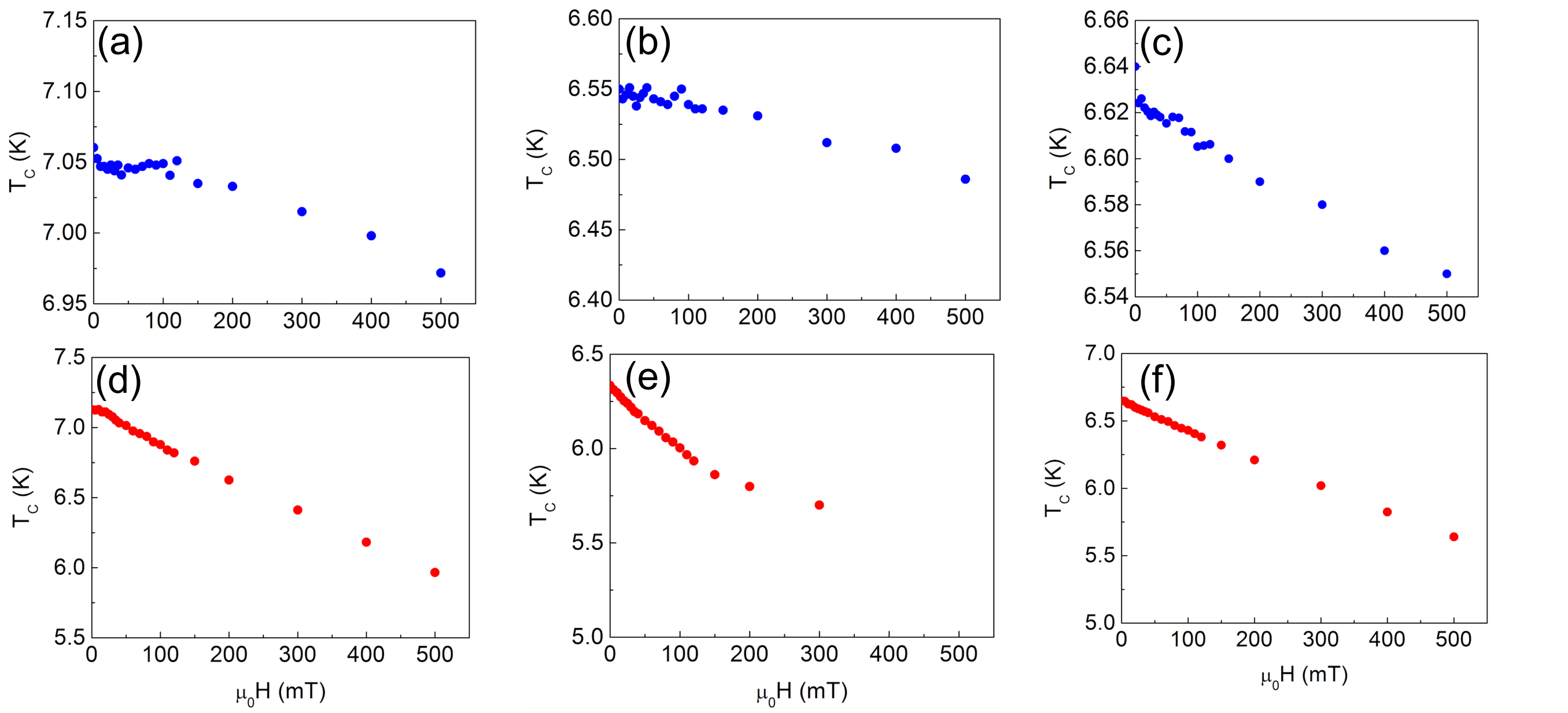

FIG. 4: Experimentally measured critical temperature $T_c$ plotted against the IP applied field $H$ for three thin-film stacks: (a) Nb(24)/Pt(2.0), (b) Nb(24)/Co(1.5)/Pt(1.5), and (c) Nb(24)/Pt(2.0)/Co(1.5)/Pt(1.5). Subfigures (d–f) show the corresponding results for an OOP applied field.

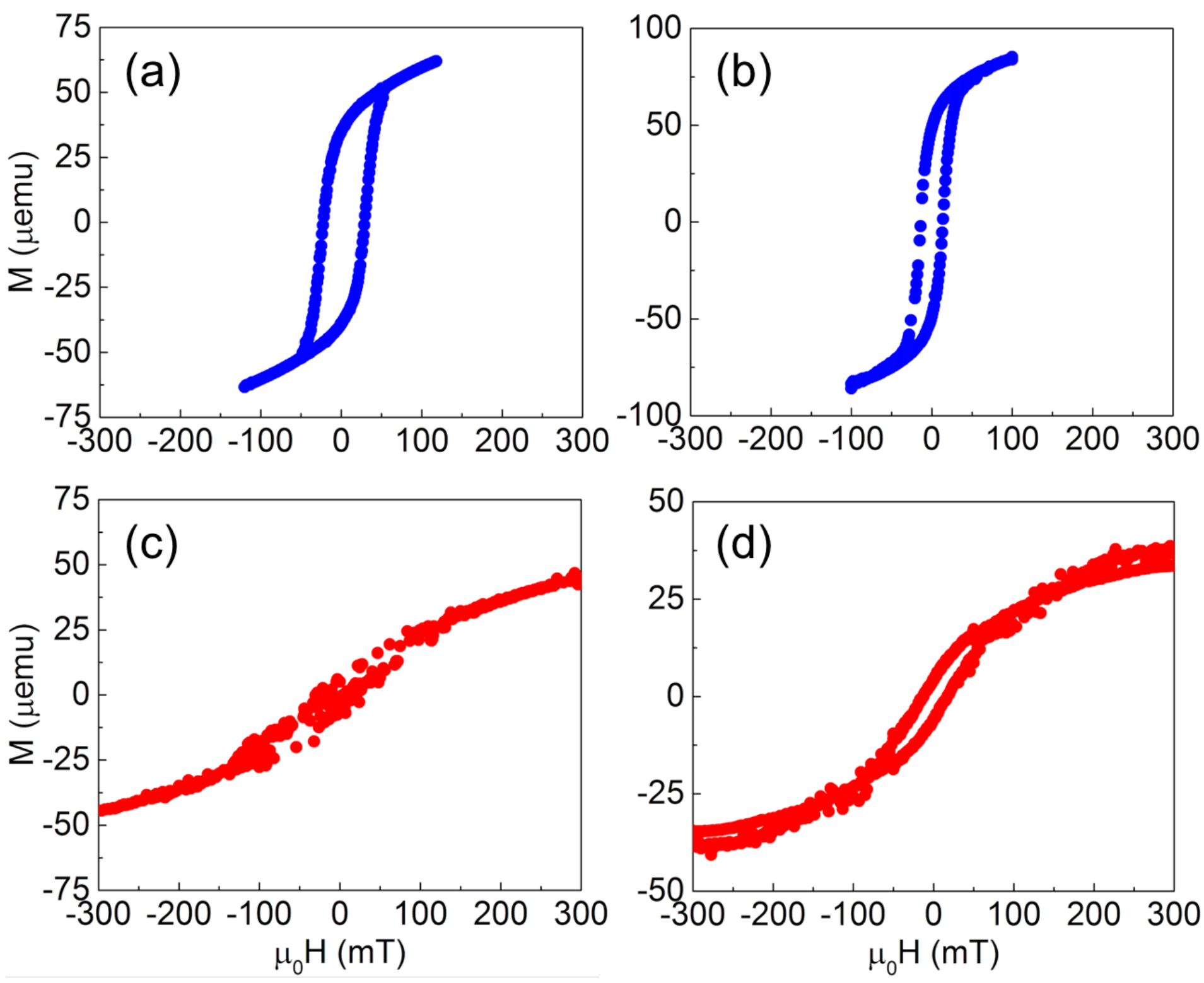

FIG. 5: $M(H)$ loops for Nb(24)/Pt(x)/Co(1.5)/Pt(1.5) samples with $x = 0.3$ (left) and $x = 1.0$ (right), for IP (top) and OOP (bottom) fields.